# Empirical Measurements of Disk Failure Rates and Error Rates

Jim Gray, Catharine van Ingen
Microsoft Research Technical Report MSR-TR-2005-166
December 2005

**Executive summary:** *The SATA advertised bit error rate of one error in 10 terabytes is frightening. We moved 2 PB through low-cost hardware and saw five disk read error events, several controller failures, and many system reboots caused by security patches. We conclude that SATA uncorrectable read errors are not yet a dominant system-fault source – they happen, but are rare compared to other problems. We also conclude that UER* (uncorrectable error rate) *is not the relevant metric for our needs. When an uncorrectable read error happens, there are typically several damaged storage blocks (and many uncorrectable read errors.) Also, some uncorrectable read errors may be masked by the operating system. The more meaningful metric for data architects is Mean Time To Data Loss (MTTDL.)*

TerraServer and other Internet properties report between 3% and 7% per year "hard" disk failures requiring drive replacement (see Table 1.) These numbers agree with anecdotes from other sites. For example, a failure rate of 8% was observed over 1000 parts years at the Internet Archive using PATA drives.

Significant error rates are also reported for disk controllers and SAN switches (see table 1).

The numbers in Table 1 are ten times the rate expected from reading the disk vendor specification sheets. This raises the question of whether other failure specifications are optimistic.

**Disk hardware specs predict an uncorrectable error in every 10 TB to 1,000 TB read.** Vendor drive specifications predict an uncorrectable bit error rate (UER) every $10^{15}$ to $10^{16}$ bits read for SCSI and $10^{13}$ to $10^{15}$ bits read for various PATA and SATA drives. But, that is just the drive error rate. Architects must add the error rate for the disk controller, the cables, the PCI bus, the memory, and the processor, so observed uncorrectable bit errors will be more frequent than the drive-level fault rate. Architects must also recognize that any of these errors can be masked by retry or other error correction strategies in the controller or operating system software

When you consider that the specified drive failure rates are more than ten times too optimistic (see Table 1), the actual uncorrectable error rate could be even worse than this one-in-10TB estimate. Data pipeline processing, data mining, and backup/restore routinely read tens of terabytes per day. Polling our sources turned up very few uncorrectable read errors – in some cases, none – so we decided to set up an experiment of our own to look for them.

| Table 1: Observed failure rates. | | | | | |
|---|---|---|---|---|---|
| **System** | **Source** | **Type** | **Part Years** | **Fails** | **Fails /Year** |
| TerraServer SAN | Barclay | SCSI 10krpm | 858 | 24 | 2.8% |
| | | controllers | 72 | 2 | 2.8% |
| | | san switch | 9 | 1 | 11.1% |
| TerraServer Brick | Barclay | SATA 7krpm | 138 | 10 | 7.2% |
| Web Property 1 | anon | SCSI 10krpm | 15,805 | 972 | 6.0% |
| | | controllers | 900 | 139 | 15.4% |
| Web Property 2 | anon | PATA 7krpm | 22,400 | 740 | 3.3% |
| | | motherboard | 3,769 | 66 | 1.7% |

Modern disks, when read sequentially, transfer about 50MB/s. At that rate, one disk can transfer $10^{14}$ bits within 4 days. So, we should be able to measure a $10^{-14}$ bit error rate. We set up JBOD SATA drives with a variety of disks and controllers on four systems (Table 2). The first system was in an office environment with no week-end air conditioning – but the system did have adequate fans. In other words, it is typical of small-office, home-office, and low-budget organizations. The remaining systems are rack mounted in an environment with full air conditioning. In other words, these systems are typical of data centers and higher-budget organizations.

We wrote a program that, for each of the disks, repeatedly did the following:
(1) Write a 10GB file with random data (a different pattern each time) computing the 64-bit checksum.
(2) Read the file computing the checksum.
(3) Write a trace record indicating the read, write, and cpu times, and whether the checksums matched.

We believe that (1) the checksum test detects almost all data errors, (2) the Windows Server 2003 event log

| Table 2: Configurations of the various test systems used to explore bit error rates. | |
|---|---|
| System 1<br>Office environment | Tyan S2882 K8S Motherboard, 2 Ghz Opteron 246, 4 GB RAM (PC 2700 with ECC) |
| | One each of the following SATA controllers: 3Ware 8400, SuperMicro "Marvell" MV8, RAIDCORE BC4452, Silicon Image Sil 3114 SATAlink |
| | 4 Western Digital 250GB  WD250JD<br>4 Maxtor 250 GB           7Y250MO |
| | Windows Server 2003 Standard with SP1 (32 bit mode). |
| System 2<br>Data center environment | Tyan S2882 K8S Motherboard, 1.8 Ghz Opteron 244, 2 GB RAM (ECC) |
| | SuperMicro "Marvell" MV8 |
| | 3 Seagate 400GB ST3400832AS |
| | Windows Server 2003 Standard with SP1 and R2 Beta 2 (32 bit mode). |
| System 3<br>Data center environment | Tyan S2882 K8S Motherboard, 1.8 Ghz Opteron 244, 2 GB RAM (ECC) |
| | SuperMicro "Marvell" MV8 |
| | 3 Seagate 400GB ST3400832AS |
| | Windows Server 2003 Standard with SP1 (32 bit mode). |
| System 3<br>Data center environment | Tyan S2882 K8S Motherboard, 1.8 Ghz Opteron 244, 2 GB RAM (ECC) |
| | SuperMicro "Marvell" MV8 |
| | 3 Seagate 400GB ST3400832AS |
| | Windows Server 2003 Standard with SP1 and R2 Beta 2 (32 bit mode). |

signals any unrecoverable bit errors on read, and (3) the SMART counters (C7 CRC errors and C8 write errors) report errors. We ran the program on the office system from May 2005 to Sept 2005 and started running the program on the data center systems in August 2005.

We ran more than 35,000 read-write cycles on the office system and over 32,000 read-write cycles on the data center systems; this corresponds to about six months of heavy drive usage on both systems. We moved more than 1.3 petabytes or 10 petabits ($10^{+16}$ bits) through the IO systems.

After running these tests for 6 months we changed the test strategy on System 1. We filled each of its 8 drives with a 100GB file. Then we ran a checksum verifier on each of the files in parallel – a cpu-bound task that runs at about 30TB per day (420 MB/s). This test detects "rotten bits" that have decayed since being written (the previous tests read bits that had just been written).

The test looks at bits that are a few months old. Thus far, the System 1 read-only test has read 756 TB and seen no errors.

These tests moved 2 petabytes and read more than 1.4 petabytes. If uncorrectable read errors are actually independent, we would expect to have seen 14 or 140 uncorrectable read errors depending on the actual UER rate. The drive specifications of UER=$10^{-14}$ suggest we should have seen 112 read errors.

To date, we observed four clear read error events and one suspected event. Each event had a slightly different signature. Each of the clear events was accompanied by one or more system error log entries as shown in Figure 1; there was no system error log entry in the suspected case. In two of the clear events and in the suspected event, the program (and the .NET runtime and Windows 2003 operating system) reported a corresponding error stack trace as shown in Figure 1.

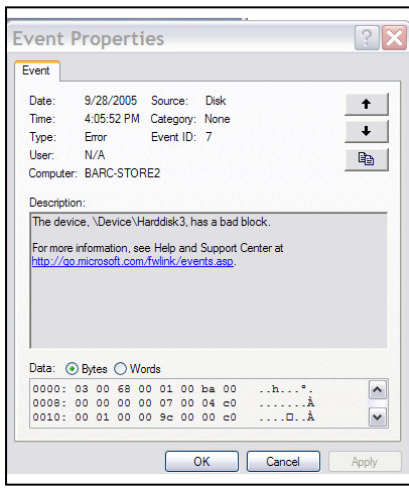

**Figure 1.** Screen shot of system error log entry associated with an uncorrectable read error (left) and error stack trace from the program receiving the error (below).

System.IO.IOException: Data error (cyclic redundancy check).
  at System.IO. Error.WinIOError(Int32 errorCode, String FullPath)
  at System.IO.FileStream.ReadCore(Byte[ ] buffer, Int32 offset, Int32 n)
  at System.IO.FileStream.Read(Byte[ ] array, Int32 offset, Int32 count)
  at System.IO.BinaryReader.FillBuffer(Int32 numBytes)
  at System.IO.BinaryReader.ReadUInt64()
  at WriteReadChecksum.ReadFile.readFile(String fileName)'

In the suspected case, the program reported an error but there were no system error log entries. At about the same time, a write error occurred on another disk attached to the same controller. Shortly after, the system crashed. The write error is likely due to a controller problem – the write failed because the disk was no longer present. Unfortunately, in this case, the system error log was lost due to corruption. We don't know whether the block mode read error was actually a block mode read error or whether the disk controller got confused and returned a questionable status code.

In two of the clear events, the program reported an error and there were bad block entries in the system error log. In one event, there were 14 error log entries with 6 different disk addresses recorded over a minute. In the other event, there were 10 error log entries with 3 different disk addresses recorded over less than a minute. The multiple entries are an artifact of how our 1 MB request is processed by Windows. Each 1 MB read request is broken into sixteen 64KB requests and queued to the disk by the operating system. When the file system receives a bad block error on a read, the read is retried 4 times. If the retries fail, the full request is then canceled. The error log entries are the result of different blocks being retried while other blocks also encounter errors before all other requests are cancelled.

In the other two clear events, there were bad block entries in the system error log but the program continued to run and reported no error. In these events, a single bad block entry was logged. The disk extent was successfully accessed when the file system retried the read – so the read error was masked by the OS retry mechanism. No error was seen by the .NET runtime or program. One of these events was accompanied by controller timeouts and occurred during patch downloading and application.

From the point of view of the programmer, we have seen 3 uncorrectable read errors. From the point of view of the operating system and the disk drive, there have been 30 uncorrectable read errors. There are actually at least 12 disk extents which encountered errors that were unrecoverable by the disk; 3 of these remained unreadable after four retries by the file system and imply data loss.

We conclude from this is that the uncorrectable read error rate as quoted by the disk manufacturers is not very useful in practice. A better metric would be mean time to data loss. We observed 3 loss events while reading 1.4 PB, This is 3 lost files. In the datacenter environment we lost 2 of 32,000 10GB files, In the office setting (System 1) we lost one 10GB file in 35,000 tries, and no 100GB files in 7,560 tries.

Read errors were not the dominant source of system outages. We experienced at least 4 controller firmware or driver errors – usually associated with a subsequent system crash or hang. Each system was updated and rebooted monthly for security patches. System version updates also caused some outages. A file system corruption and a disk timeout causing it to go offline co-occurred with some of these installs.

These system restarts did not cause data loss – just service interruption. The only data loss was the 3 uncorrectable read errors mentioned above and the corruption and then loss of the System1\C: system disk. The corruption was repaired by the operating system utility (chkdisk) and lost binaries were recovered (using the Recovery Console). A month later that disk failed completely and was replaced. There were no disk failures among our test drives (System1\C: was not one of the test drives). So, one could argue that we had data loss from one drive failure and from 3 uncorrectable read errors.

We continue these experiments; this is just a status report. We also continue to work with Internet-scale properties who are tracking their storage failure rates.

**Acknowledgements**

We are grateful to Dave Anderson, Tom Barclay, Bruce Baumgart, Molly Brown, Garret Buban, William Casperson, Andrew Cencini, Leo Cazares, Wyman Chong, Wayne Flagg, Bob Fitzgerald, Dustin Fraser, Jeffrey Goldner, Darryl Havens, Andrew Kadatch, Matthew Kerner, Mark Licata, Mark Manasse, Jeff Meng, Ravisankar Pudipeddi, Gilad Sade, David Schiffrin, Sunita Shrivastava, Phil Smoot, Dan Stevenson, Dan Teodosiu, Balaji Thiagarajan, Jeremy Winter, Mark Zbikowski, and our other anonymous friends in the disk industry who helped us with data and helped us understand these issues.